\documentstyle[aps]{revtex}

\begin{document}
\flushbottom

\def\thepage{\roman{page}}
\title{\vspace*{1.5in}Gravitational waves, baryogenesis, and dark matter from
primordial black holes.} 

\author{A.D. Dolgov}

\address{INFN section of Ferrara, Via del Paradiso 12,
44100 Ferrara, Italy;\\ ITEP, Bol.Cheremushkinskaya 25, Moscow 117259, Russia.}

\author{P.D. Naselsky}

\address{Theoretical Astrophysics Center, Juliane Maries Vej
30, 2100 Copenhagen, {\O} Denmark;\\ Rostov State University, Zorge 5, 344090
Rostov-Don, Russia.}

\author{I.D. Novikov}

\address{Theoretical Astrophysics Center, Juliane Maries Vej
30, 2100 Copenhagen, {\O} Denmark;\\ Astro-Space Center of Lebedev Physical Institute, Profsoyuznaya
84/32, Moscow, Russia;\\ NORDITA, Blegdamsvej 17, DK- 2100, Copenhagen, Denmark.}

\maketitle

\setlength{\baselineskip}{0.20in}

\newcommand{\nc}{\newcommand}
\newcommand{\beq}{\begin{equation}}
\newcommand{\eeq}{\end{equation}}
\newcommand{\be}{\begin{eqnarray}}
\newcommand{\ee}{\end{eqnarray}}
\newcommand{\num}{\nu_\mu}
\newcommand{\nue}{\nu_e}
\newcommand{\nut}{\nu_\tau}
\newcommand{\nus}{\nu_s}
\newcommand{\mnus}{M_s}
\newcommand{\taus}{\tau_{\nu_s}}
\newcommand{\nnt}{n_{\nu_\tau}}
\newcommand{\rnt}{\rho_{\nu_\tau}}
\newcommand{\mnt}{m_{\nu_\tau}}
\newcommand{\tnt}{\tau_{\nu_\tau}}
\newcommand{\bi}{\bibitem}
\newcommand{\rar}{\rightarrow}
\newcommand{\lar}{\leftarrow}
\newcommand{\lrar}{\leftrightarrow}
\newcommand{\dm}{\delta m^2}
\newcommand{\mpl}{m_{Pl}}
\newcommand{\mbh}{M_{BH}}
\newcommand{\nbh}{n_{BH}}

\makeatletter
\def\alt{\mathrel{\mathpalette\vereq<}}
\def\vereq#1#2{\lower3pt\vbox{\baselineskip1.5pt \lineskip1.5pt
\ialign{$\m@th#1\hfill##\hfil$\crcr#2\crcr\sim\crcr}}}
\def\agt{\mathrel{\mathpalette\vereq>}}

\newcommand{\eq}{{\rm eq}}
\newcommand{\tot}{{\rm tot}}
\newcommand{\M}{{\rm M}}
\newcommand{\coll}{{\rm coll}}
\newcommand{\ann}{{\rm ann}}
\makeatother

\begin{abstract}
We discuss the hypotheses that the cosmological baryon asymmetry and 
entropy were produced in the early Universe by the primordial black 
hole (PBHs) evaporation. It is assumed that the Planck mass remnants 
of the PBHs evaporation survive to the present day and form the cold
dark matter of the Universe. It is shown that high energy gravitons 
emitted by the PBHs form new relic with typical energy of particles 
about 100 eV. The energy density of gravitons at the present time of 
the cosmological expansion is equal to 0.3 - 1 \% of the density of 
the cosmic microwave background radiation.

\end{abstract}

\section{Introduction}

After the pioneering papers~\cite{novikov67,hawking71} 
several possible mechanisms of primordial black hole production 
were described in the literature, for a review see e.g. the 
book~\cite{frolov}. Normally such black holes are sufficiently light
so that they evaporated by the Hawking mechanism~\cite{hawking74}
leaving behind relativistic products of evaporation that subsequently
thermalized with the primeval plasma. The ultimate fate of these 
evaporating black holes is unknown, the calculations are made in
quasiclassical approximation when effects of quantum gravity are not
essential. The latter must come into play when the mass of black holes 
drops down to the Planck magnitude, $\mbh = \mpl = 1.22 \times 10^{19}$ GeV.
Our present knowledge does not permit  to conclude whether the 
evaporation stops at  $\mbh \sim \mpl$ and a stable remnant would survive
or black hole would disappear completely decaying into usual elementary
particles. The resolution of this problem depends upon the ultraviolet
behaviour of quantum gravity and if the latter is asymptotically free
the hypothesis of stable remnants of evaporation with the Planck mass
looks quite probable. We will assume that it is true throughout this paper.

We will consider here the following scenario. Primordial black holes
have been produced in the early universe by one or other mechanism 
which we do not specify. Their
life-time with respect to evaporation is assumed to be sufficiently long 
so that the universe became dominated by the gas of non-relativistic 
black holes prior to their decays. After black hole evaporation a new 
relativistic epoch is created by the thermalized products of the
evaporation. Baryon asymmetry of the universe can be created at this
stage through the mechanism outlined by Hawking~\cite{hawking74} and
Zeldovich~\cite{zeldovich76}. The Planck mass remnants of the evaporation
survive to the present day and form the cold dark matter of the 
universe. Gravitons emitted by the black holes did not thermalize and
at the present time create a high frequency gravitational radiation
with a rather large cosmological energy density. Collisions of the
Planck mass mini-black-holes may create  a black holes with a masse higher then a
stable remnant. This leads to the evaporation and may produce rare events of high 
energy cosmic rays.

\section{Thermal history of the universe early dominated by primordial 
black holes.\label{pbh}}

At the end of inflation the universe became radiation dominated by mostly
light products of the inflaton decay. However some fraction of matter
could form primordial black holes~\cite{ivanov94}-\cite{green00}.
Let us denote this initial fraction as 
$\epsilon_{i} =\rho_{BH}/ \rho_{tot}$. For simplicyty we suppose that
all primordial black holes have the same mass $\mbh$. The relative role of 
non-relativistic black hole matter was increasing in the course of 
expansion and when the scale factor rose by $1/\epsilon_{BH}$ black
holes became the dominant cosmological matter. This should happen before
their evaporation so the relative initial mass fraction of black holes 
must satisfy the condition:
\be
\epsilon_i > \sqrt{{t_i \over \tau}} \approx 
0.03 \left({\mpl \over \mbh}\right)^{3/2}\, {\mpl \over T_i}
\label{epsi}
\ee
where $T_i$ is the initial plasma temperature after the end of inflation
and $t_i$ is the ``initial'' time related to the initial energy density
through
\be
\rho_i = {3 \over 32 \pi} { \mpl^2 \over t_i^2} = 
{\pi^2 \over 30} g_* T_i^4
\label{rhoi}
\ee
The factor $g_* \sim 100$ is the number particle species in thermalized 
plasma. The black hole life time, $\tau$, is given by the 
expression~\cite{page76}:
\be
\tau \approx 3\times 10^3 N_{eff}^{-1} \mbh^3 \mpl^{-4}
\label{tau}
\ee 
where $N_{eff} \sim 100$ is the effective number of particle species with
masses smaller than the black hole temperature
\be
T_{BH} = { \mpl^2 \over 8\pi \mbh}
\label{tbh}
\ee

As we have found below  $\mbh \sim 10^{11}\mpl$.
Since normally $\epsilon_i$ should be much smaller than unity, the validity
of the condition (\ref{epsi}) implies that the (re)heating temperature
after inflation should be sufficiently high, $T_i \gg 10^{-18} \mpl$.
This is rather a mild condition.

As we mention above we assume that the mass distribution of black holes is strongly peaked
near some mass value $\mbh=M_0$ to be determined in what follows. In
particular the model of an early black hole formation suggested in
ref.~\cite{dolgov93} predicts the log-normal mass distribution,
$dN /dM \sim \exp (-\gamma \log^2 M/M_0)$, that may be a justification of
the above assumption. We will also adopt the instant decay approximation
for the black hole evaporation. According to that, all black holes decayed
instantly at the same time $t=\tau$ and this led to an instant change
of the matter dominated expansion to the relativistic one. The cosmological
energy density at that time was
\be
\rho_d = {\mpl^2 \over 6\pi \tau^2} = 
6\times 10^{-5}\,\left({N_{eff} \over 100}\right)^2 {\mpl^{10}\over 
\mbh^6}  
\label{rhod}
\ee
The number density of the black holes can be found from assumption that 
all or predominant part of dark matter in the universe consists of the
Planck mass mini-black holes that are the remnants of the evaporation of
their large parents that took place at $t=\tau$. It means in particular
that the comoving number density of the parent black holes at the moment 
of their evaporation is equal to the comoving number density of their
Planck mass remnants at the present time. The latter is equal to:
\be
n_{BH}^{(0)} = {\Omega_{DM}\over \mpl} = 
10^{-25} \,{\rm cm}^{-3}\,\left({\Omega_{DM} \over 0.3}\right)
\left({h \over 0.65}\right)^2
\label{nbh0}
\ee
where $\Omega_{DM} =\rho_{DM}/\rho_c$ is the relative mass fraction of
the dark matter now and $h = H/ 100\,{\rm km/sec/Mpc}$.

We can estimate the characteristic mass that black holes should have
in order to satisfy the conditions considered above from the following
considerations. The number density of the black holes at the moment of
their decay is given by
\be
n_{BH}^{(d)} = (z_d+1)^3\, n_{BH}^{(0)}
\label{nbhd}
\ee
where $z_d = a_0/a_d$ is the red-shift at the moment of the decay and
$a$ is the cosmological scale factor. The cosmological energy density
at the moment of the decay is given by eq.~(\ref{rhod}) on one hand side 
and by the evident expression 
\be
\rho_d = \mbh n_{BH}^{(d)},
\label{rhodm}
\ee
on the other. For the first approximate estimate we assume that the 
energy density of relativistic matter decrease as $1/a^4$. This is true 
if no entropy is released in the plasma (e.g. by annihilation of
massive particles). In this case $ \rho_d / \rho_0 = (z_d +1)^4$, 
where $\rho_0 = 0.44 {\rm eV/cm}^3$ is the energy density of 
relativistic matter now (photons plus three neutrino species). From
this approximate relation and equations~(\ref{nbhd},\ref{rhodm}) 
follows 
\be
\mbh n_{BH}^{(0)} = \rho_d^{1/4} \rho_0^{3/4} 
\label{mbhnbh}
\ee
and after a simple algebra we obtain
\be
\mbh = 7.6 \times 10^{10}\,\mpl
\left({N_{eff}\over 100}\right)^{1/5}
\left({\Omega_{DM}\over 0.3 }\right)^{-2/5}
\left({h\over 0.65}\right)^{-4/5}
\label{mbh}
\ee  

A more accurate result for $\mbh$ can be obtained if entropy release is
taken into account.
Assuming that the thermalization of the evaporation products is fast on
the cosmological time scale we can obtain the plasma temperature immediately
after the black hole decays from the equality:
\be
\rho_d = {\pi^2 \over 30} \, g_*^{(d)} T_d^4
\label{td}
\ee
where $g_*^{(d)} \sim 100$ is the number of relativistic degrees of freedom 
in the cosmic plasma at $T=T_d$ and $\rho_d$ is given by eq.~(\ref{rhod}). 
For the case of adiabatic expansion (which 
is true if there are no strong first order phase transitions) the entropy in 
coming volume is conserved, so that
\be
g_* a^3 T^3 =const
\label{gat}
\ee
This condition is fulfilled if the species are in thermal equilibrium and
if chemical potentials are vanishingly small (see e.g. the 
book~\cite{weinberg}). Thus we may expect that down to temperatures of the
order of a few MeV when neutrinos, electron/positrons, and photons were
in thermal equilibrium, the red-shift and temperature evolution satisfy
the relation:
\be
g_*^{(d)}\, T_d^3\, a_d^3 =g_*^{(W)}\, T_W^3\, a_W^3 
\label{gta}
\ee
where $g_*^{(W)}=10.75$ is the number of species at 
$T\sim T_W \sim (\rm a\,\,\, few\,\, MeV)$. Below $T_W$ neutrinos drop
out of thermal contact with the rest of the plasma and their temperature
evolves simply as $T_\nu \sim 1/a$. Thus the red-shift corresponding to 
the decay of the black holes is given by
\be
z_d +1 \equiv \left({ a_0 \over a_d}\right) = 
\left({g_*^{(d)} \over g_*^{(W)}}\right)^{1/3} \,{T_d \over T^{(0)}_\nu}
\label{zd}
\ee
where $T^{(0)}_\nu$ is the present day temperature of the cosmic neutrino
background; in the standard cosmological model it is
$T^{(0)}_\nu = (4/11)^{1/3} T_\gamma^{(0)} = 1.9\,{\rm K} = 
1.66\times 10^{-4} $ eV.

Using the presented above expressions we obtain for the mass of
the black holes 
\be
\mbh = 5.7\times 10^{10}\,\mpl\left({100\over g_*^{(d)}}\right)^{1/10}
\left({N_{eff}\over 100}\right)^{1/5}
\left({\Omega_{DM}\over 0.3 }\right)^{-2/5}
\left({h\over 0.65}\right)^{-4/5}
\label{mbh2}
\ee 
The temperature of plasma after the products of the evaporation were 
thermalized can be found from eq.~(\ref{rhod},\ref{td}),
$T_d \approx 30$ GeV, while the black hole temperature according
to eq.~(\ref{tbh}) is 
\be
T_{BH} = 8.4\times 10^6\,\,{\rm GeV}
\label{tbhnum}
\ee

\section{Thermalization and gravitational radiation. \label{gravrad}}

All particles with the mass smaller than $3T_{BH}$ are efficiently 
produced in the process of evaporation. Their initial spectrum was
thermal on the horizon but in the process of their propagation in the
gravitational field of the parent black hole the spectrum was somewhat
distorted. This distortion was calculated in ref.~\cite{page76} but
in our approximate estimates we neglect this distortion. Since
the black hole temperature is inversely proportional to $\mbh$ the 
evaporation has an explosive character and one may assume that all
black holes disappeared instantly at $t=\tau$. The average number density
of the produced particles was 
\be
n_{part} \sim  {\rho_{BH} \over 3T_{BH}} = {4\over 3} {\mbh \over \tau^2}
\label{npart}  
\ee
The rate of thermalization is equal to 
\be
\dot n_{part} /n_{part} = \sigma n_{part} N_{eff}
\label{dotn}
\ee
where $\sigma$ is the the particle interaction cross-section; it is
typically equal to 
$\sigma \approx \alpha^2 /E^2 \approx \alpha^2/T_{BH}^2 $ and 
$\alpha$ is the characteristic value of the coupling constant. At
high temperatures or energies given by eq.~(\ref{tbh}) its value
is about 1/50, approximately twice bigger than at low energies due
to logarithmic renormalization.

Thermal equilibrium is established when this rate is larger than the 
universe expansion rate given by the running value of the Hubble
parameter. To the moment of the decay of the black holes the latter
is approximately equal to the inverse life-time of the black holes, 
$H \approx 1/\tau$. The condition of equilibrium at $t=\tau$ reads:
\be
{ 2^8\times10^4\, \pi^2 \alpha^2 \over 3^3\times 10^3}\,
\left({N_{eff} \over 100 }\right)^2 >1
\label{equil}
\ee
However for $\alpha = 0.02$ and $N_{eff}=100$ the l.h.s. of this
inequality is only 0.37, so the equilibrium is not immediately
established. However it would be established a little later at
$t>\tau$. This would not lead to any significant change of the 
results presented in the previous section.

Of all the particles produced by the evaporation gravitons were 
definitely not thermalized. Their interaction with the cosmic plasma
is too weak so that they would arrive to our time undisturbed and
only red-shifted. Initially gravitons carry about 1\% of the total 
evaporated energy and to the present day this fraction decreases  
approximately by factor 3 due to entropy release. 
Thus at the present time the energy density of gravitons is about
0.1 eV/cm$^3$ and their average energy is 
$3\,T_{BH}/(z_d+1) \approx 10^2$ eV or the frequency 
(a few)$\times 10^{16}$ Hz. Unfortunately this frequency range is
far outside the sensitivity of the existing and proposed detectors of
gravitational radiation.

\section{ Baryon asymmetry.}

The reheating temperature after black hole evaporation is quite low, 
about 30 GeV, (see the end of sec. 1) and generation of cosmological
baryon asymmetry could meet some difficulties. There are two possible
ways out. Baryon asymmetry might be generated prior thermalization of
the plasma when the particle energies were quite high, of the order of
black hole temperature. Another, and more attractive for the considered
model, way is to generate baryon asymmetry in the process of evaporation
itself. The possibility that black hole evaporation could create an 
excess of baryon over anti-baryons was noticed in 
refs.~\cite{hawking74,zeldovich76} and detailed calculations were done
in papers~\cite{dolgov80}. The idea was criticised in 
ref.~\cite{toussaint79} because the black hole evaporation produces 
a thermal equilibrium state and in thermal equilibrium no asymmetry
between particles and anti-particles can be developed. This is indeed
true for particle emission at the black hole horizon, however after
that they interact between themselves and with the gravitational field of
the parent black hole. These interactions cold permit to develops a
considerable asymmetry. A concrete model discussed in 
refs.~\cite{zeldovich76,dolgov80} was the following. A heavy unstable 
neutral particle $\chi$, emitted at the black hole horizon, decayed in the 
gravitational field of the black hole into two channels:

\be
\chi \rar B_1 \bar B_2, \,\,\, B_2 \bar B_1,
\label{chib}
\ee
where $B_1$ and $B_2$ are some particles with an equal and non-zero
baryonic charges and with different masses. For example the particle
$\chi$ could be a neutralino decaying into a heavy squark, $B_1$ and 
a much lighter quark, $B_2$. 
Due to C and CP breaking the probability of these two decay channels
could be different, so for example, more quarks are produced by the
decay than anti-quarks. The probability of the inverse capture of the 
produced particles by the parent black hole is larger for heavier
particles. Thus more squarks and anti-squarks are back-captured by the
black hole. As a result a non-zero flux of baryonic charge into 
external space would be created. 

It is worth noting that the underlying
microscopic theory may conserve baryonic charge. An effective 
non-conservation of baryons are induced by the interaction with
disappearing black holes. In this model anti-baryons are accumulated 
inside evaporating black hole that disappears into nothing or, as we 
assume here, into astrophysically light Planck mass remnants with a 
gigantic anti-baryonic charge. Thus according to the discussed here
scenario of baryogenesis, the universe has net zero baryonic charge but 
it is carried by different particles: baryonic charge is carried by the 
usual baryonic matter, protons and neutrons, while anti-baryonic  charge 
is concealed inside the Planck mass mini-black holes. It is similar
to distribution of electric charge: positive charge is carried by protons,
negative charge is carried by electrons and the total electric charge of
the universe is identically zero. However in contrast to the baryon
asymmetry, no electric asymmetry
can be generated by the described here mechanism because
there is the long range electromagnetic interaction associated with 
electric charge, so that black holes have electric hairs, while we
believe that baryonic hairs are absent. The upper limit on the possible
long-range forces induced by baryonic charge has been found from the
validity of equivalence principle in ref.~\cite{lee56} (for a recent
review see~\cite{dolgov99}): $\alpha_B < 10^{-46}$, to be compared
with the coupling strength of electromagnetic interaction,
$\alpha_{em} = 1/137$.

Rather lengthy calculations of ref.~\cite{dolgov80} give the following
result for the cosmological baryon asymmetry (entropy per baryon)
\be
\beta \approx 0.1 N_{eff}^{-1/4} {\rho_{BH} \over \rho_{tot}}\,
{\Delta \Gamma_\chi \over m_\chi} \left({\mpl \over \mbh}\right)^{1/2}
\label{beta}
\ee
where $\Delta \Gamma_\chi = \Gamma_\chi (\chi\rar B_1 \bar B_2)-
\Gamma_\chi (\chi\rar B_2 \bar B_1)$ is the difference between the
widths of the two charge conjugated decay channels. Since the effects 
of C and CP violation can be observable only if the re-scattering of the 
decay products is taken into account, the difference $\Delta \Gamma_\chi$
is non-vanishing only in the second order in the coupling constant,
that in the considered energy range is about 1/50. Thus,
${\Delta \Gamma_\chi  m_\chi} \approx \alpha^2 \approx 4\times  10^{-4}$.
This result would be somewhat larger if several decay channels of this kind
are open. Since the black holes dominated the cosmological energy density at
the moment of their evaporation, the ratio 
${\rho_{BH} / \rho_{tot}}$ is close to unity and we obtain 
$\beta \approx 10^{-10}$. This is very close to the observed value.

\section{Conclusion}

We have examined how relatively small PBHs could create cosmologically
interesting cold dark matter,  baryon asymmetry, and the entropy
of the Universe. The crucial assumption for creation of the cosmological 
dark matter is connected with the  
the final stage of the PBHs evaporation when the black hole masses 
reach the Planck limit $\mbh\approx \mpl$. It is assumed usually that
the evaporation does not stop at $\mbh\sim \mpl$ so that the black hole
disappears completely due to the Hawking process, though strongly
modified by quantum gravity effects. However the physical reason for such
extrapolation of the quasiclassical Hawking's analysis is unclear and it
is possible that the evaporation switches off at the Planck scale. We
cannot prove that this conjecture is true but it is very attractive
and has at least the same right to exist as the assumption of the
complete disappearance of evaporating black holes. To resolve the 
problem one has to understand the ultraviolet behaviour of quantum
gravity.

We have shown that the products of the evaporation of light black holes
with $\mbh\approx 10^{6}$ g could have generated the 3 K background
radiation in initially warm or cold Universe after inflation. This idea 
is not new, but a new point is a possible cold dark matter formation by 
this mechanism. We have predicted the spectrum and characteristic energy 
of a new possible relic of the entropy epoch formation, namely, the 
energetic gravitons from PBHs evaporation with $E \sim 100$ eV. An 
observation of such gravitons would be a critical test of our hypothesis.
Some argument in favour of the presented model is that the baryon
asymmetry of the universe that could be created by evaporating PBHs,
even if baryonic charge is strictly conserved, is of the right 
magnitude.

\section*{Acknowledgement}

A.~Dolgov is grateful to the Theory Division of CERN for the
hospitality during the period when this work was completed.

This investigation was supported in part by a grant RFBR 17625, by a grant INTAS 97-1192, by
the Danish Natural Science Research Council through grant No 9701841 and by 
Danmarks Grundforskningsfond through its support for the establishment of the
Theoretical Astrophysics Center.

\end{document}